# Instantaneous mapping of liquid crystal orientation using a polychromatic polarizing microscope

Mojtaba Rajabi [1,2], Oleg Lavrentovich[1,2,3] and Michael Shribak[4,*]

[1]Advanced Materials & Liquid Crystal Institute, Kent State University, Kent, OH 44242, USA
[2]Department of Physics, Kent State University, Kent, OH 44242, USA
[3]Materials Science Graduate Program, Kent State University, Kent, OH 44242, USA
[4]Marine Biological Laboratory, Woods Hole, MA, 02543, USA

*Email: mshribak@gmail.com

Polarizing microscopy brought about many advancements in the science of liquid crystals and other soft materials, including those of biological origin. Recent developments in optics and computer-based analysis enabled a new generation of quantitative polarizing microscopy which produces spatial maps of the optic axis. Unfortunately, most of the available approaches require a long acquisition time of multiple images which are then analyzed to produce the map. We describe a polychromatic polarizing microscope, which allows one to map the patterns of the optical axis in a single-shot exposure, thus enabling a fast temporal resolution. We present a comparative analysis of the new microscope with alternative techniques such as a conventional polarizing optical microscope and MicroImager of Hinds Instruments.

## INTRODUCTION

A polarizing microscope is the most common tool to characterize materials, including liquid crystals (LCs). Invented by Giovanni Amici almost two centuries ago [1], its principal structure, function, and analytical capabilities have not changed much since then [2, 3]. Progress in the developments of computers, digital cameras, and optical components with electrically controlled polarization, such as electro-optical and magneto-optical crystals, liquid crystal cells, etc., modernized polarizing microscopy and made it quantitative.

A contemporary quantitative polarizing microscope maps the two-dimensional (2D) phase retardance of the sample. The map is produced by two optical components. The first component called a polarization state generator (PSG) creates a beam with certain switched polarization states. The second optical component, placed between a specimen and a digital camera, is a circular or rotatable linear analyzer. The analyzer transforms the phase difference between two orthogonal polarization modes into the transmitted intensity map. The earliest 2D retardance mapping techniques were reported in 1992 by Noguchi et al. [4] and Otani et al. [5]. Noguchi used a PSG with a rotated quarter-wave plate and the left circular analyzer. Otani built his system with a PSG comprising a rotatable variable retarder, such as a Babinet–Soleil compensator, and a rotatable linear analyzer. At the same time, Shribak proposed a new device for measuring phase retardance of reflective specimens, which employs a PSG with a rotatable polarizing beam-splitter [6-8]. This PSG works in both transmitted and reflected beams simultaneously. It was used by Shribak et al. to build a return-path setup for mapping 2D birefringence distribution [9].

In 1967, Yamaguchi and Hasunuma [10,11] proposed a non-mechanical PSG formed by two variable retarders with optical axes oriented at 45º with respect to each other. In 1994,

Oldenbourg and Mei used this approach to build a PSG with two LC cells and called it the universal compensator. The phase retardance of the LC cells could be controlled by the applied electric field, which tilts the optical axis (in an LC, the optical axis is also the director of molecular orientation). Similarly, to the Noguchi's scheme [4], they combined the universal compensator with a left circular analyzer and named the new microscope a Pol-Scope [12, 13]. Later Shribak proposed several improved algorithms, which increased sensitivity and reduced the acquisition time of the 2D retardance mapping [14-17]. These algorithms were used in LC-polscopes manufactured by former CRi Inc. (Woburn, MA), under the names Oosight and Abrio. Currently, Oosight is available from Hamilton Thorne (www.hamiltonthorne.com). Shribak also developed a polarizing microscope with a single LC cell (SLC-polscope) [18]. The SLC-polscope offers the same benefits as the CRi's LC-polscope, but it is simpler, faster, and less expensive.

The major drawback of the above-mentioned techniques is the sequential acquisition of several raw images taken with different retardance settings of the PSG, which are then processed to compute a 2D birefringence map. Usually, the set of images needed to reconstruct the map of retardance is acquired within 1 s, which is too slow to study many dynamic processes. To overcome this issue, Shribak et al. developed the real-time polscope, which simultaneously captures four raw polarization images [19, 20]. The imaging beam is divided by a special beam-splitter equipped with four elliptical analyzers. The real-time polscope does not suffer from artifacts caused by the movements of the specimen. However, it requires a complicated image alignment, which creates difficulty in practical use. It is also quite expensive.

In 2009 Shribak invented a new type of polarizing microscope called the polychromatic polscope [21, 22] equipped with a polychromatic polarizing module (PPM), which consists of a polychromatic PSG and an achromatic circular analyzer. Even at very low retardance, PPM produces a full hue-saturation-brightness (HSB) color spectrum in birefringent materials. Unlike the conventional polarizing microscopy, where interference colors measure the retardance according to the Michel-Levy chart [23], the HSB hues in PPM depend on the orientation of the slow axis orientation with respect to a preset "zero" direction (usually oriented along the East-West axis of a microscope's stage). Rotation of the slow axis continuously changes the hues, repeating them after every 180°; the state of extinction is never achieved. The polychromatic polarizing microscope allows one to see a colored polarization image by the eye and capture the picture instantly, in real-time. Therefore, it enables one to achieve a high temporal resolution, limited only by the acquisition time of the camera or the duration of the light pulse. It provides sharp images of fast-moving, low-birefringent structures and makes visible rapid processes accompanied by birefringence changes. The low-birefringence condition means that the optical retardance should be less than about 250 nm in order to avoid the Newton interference colors that would complicate the analysis.  In combination with a pulsed light source, the new microscope should make it possible to visualize dynamic effects in birefringent materials, such as the development of nerve pulses, shock wave propagation, flows of liquid crystals, etc.

In this work, we describe a new polarizing microscopy technique, which is based on the vector interference of polarized light in which the full spectrum colors are created at the retardance of several nanometers, with the hue determined by the orientation of the birefringent structure. The approach allows one to significantly shorten the acquisition time. We illustrate the principle by exploring patterned director fields and dynamic textures of sheared liquid crystals

# 1. Polychromatic polarizing module

A typical optical scheme of a PPM, comprising a polychromatic PSG and an achromatic circular analyzer, is shown in Fig.1. The PPM is available from Marine Biological Laboratory (Woods Hole, USA) and can be manufactured per request (https://www.shribaklab.com/). The module could be added to any standard polarizing microscope equipped with a white light source and a color digital camera.

The PSG produces polarized light with the polarization ellipse orientation determined by the wavelength. A set of ellipses corresponding to different wavelengths is called a spectral polarization fan. An example of the fan with the right polarization ellipses for the wavelengths in the range 440-660 nm is illustrated by the callout in Fig.1. The major axis of the red polarization ellipse (λ=660 nm) makes 45º with the horizontal axis. The major axis of the orange polarization ellipse (λ=609nm), yellow polarization ellipse (λ=566nm), green polarization ellipse (λ=528nm), cyan polarization ellipse (λ=495nm), and blue polarization ellipse (λ=466nm) are oriented at 75º, 105º, 135º, 165º and 15º to the horizontal axis, respectively. All polarization ellipses have the same ellipticity angle ε=40º. The ellipticity angle of a polarization ellipse is defined as the arctangent of the ratio of the minor to major axes. Positive and negative values of the ellipticity correspond to right-handed and left-handed polarizations, respectively [24].

When the major axis of the polarization ellipse is at 45º or 135º to the slow axis of a birefringent specimen, the intensity of light transmitted through the PSG, sample of a phase retardance d and the achromatic left circular analyzer is [18]:

$$I = I_0 \, sin^2(45º - \varepsilon \pm \delta/2), \tag{1}$$

where $I_0$ and ε are the intensity and the ellipticity of the illumination beam, respectively, and the sign "+" is taken for 45º and "-" for 135º. If the specimen under investigation is not birefringent, then the polarization of a transmitted white light beam does not change. The left circular analyzer will transmit 0.0075 intensity of all wavelengths evenly, as follows from Eq.1 for ε=40º and d =0; the output beam will remain white. If the object is birefringent, then it modifies the spectral polarization fan.

For example, consider a particle with retardation 15 nm; its phase retardation is 8º and 10º at wavelengths 660 nm and 528 nm, respectively. If the particle's slow axis is oriented at 0º and the ellipticity of the illumination beam is ε=40º then, according to Eq.(1), the transmission of the red component is 0.024, which is 3 times the transmission of the non-birefringent background. The transmission of the green component will be zero. The birefringent particle will appear predominantly red. If the particle or the spectral polarization fan is rotated by 90º, the particle image changes to a complementary green color. One can consider this device as a spectral Brace-Kohler compensator, which rotates polarization ellipse spectrally rather than mechanically.

The spectral fan of polarization ellipses could be assembled from a rotatable linear polarizer, achromatic quarter-wave plate (AQWP), and optically active waveplate (OAWP), Fig.1. The polarizer and AQWP produce a polarization ellipse with the major axis parallel to the slow axis of AQWP [24]. The waveplate OAWP is cut perpendicularly to the optical axis of a uniaxial gyrotropic crystal, such as quartz [22,25]. When the polarized light propagates along the optical axis of a gyrotropic crystal, the polarization ellipse rotates by some angle. The rotation angle is linearly proportional to the OAWP thickness and inversely proportional to the wavelength. The thickness of the waveplate is about 8 mm. The eigen polarizations of OAWP are circular, and the polarization rotation angle equals half of the phase shift between the eigen polarizations.

In principle, the ellipticity could be introduced by tilting the waveplate OAWP by a small angle, say, 10º [25]. Then the waveplate OAWP will work simultaneously as a Berek compensator [3] and a polarization rotator. In this case, the eigen polarizations of waveplate OAWP are elliptical, and the ellipticity decreases if the incidence angle increases. However, the analytical formulae for the gyrotropic waveplate are cumbersome [25], while a mathematical description of PPM with a tilted OAWP has not been developed yet.

The achromatic left circular analyzer combines an achromatic quarter-wave plate (AQWP) with a slow axis oriented at 45º and a high-quality linear polarizer (analyzer) with polarization direction at 0º. Usually, achromatic quarter-wave retarders use a combination of two or three retarders of materials having different birefringence dispersion to minimize deviation from quarter-wave retardance across a broad spectral range [26, 27]. The most common achromatic quarter-wave plates combine quartz and magnesium fluoride.

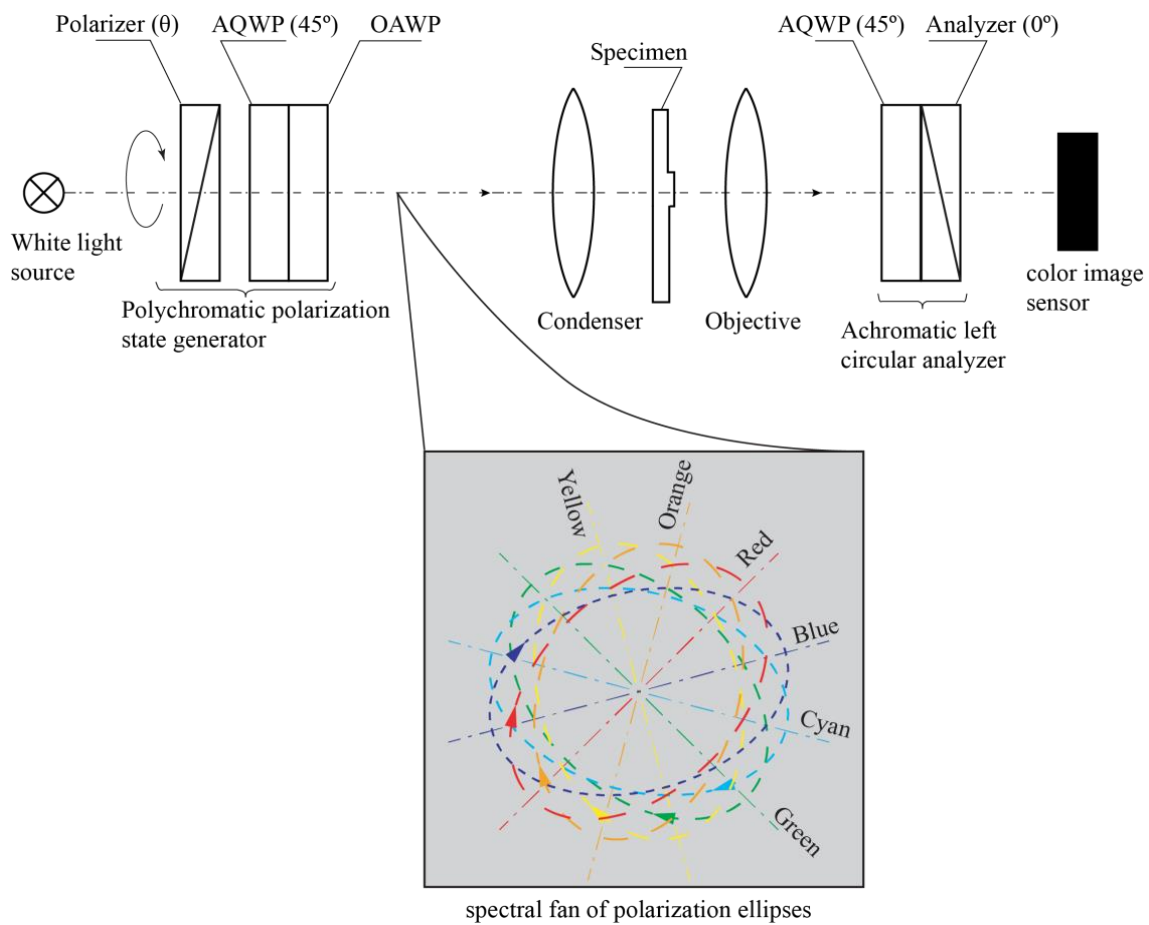

Fig. 1. Schematic of microscope with a polychromatic polarizing module (PPM) that consists of a polychromatic PSG and an achromatic left circular analyzer. OAWP is an optically active waveplate; AQWPs are achromatic quarter-wave plates with slow axes at 45º. The polarization direction of a linear analyzer is 0º (horizontal).

To illustrate the utility of PPM polarizing microscopy, we install the PPM in an upright microscope Olympus BX51, equipped with a 100W halogen lamp U-LH100-3 and a Basler PowerPack acA1920-155um digital color camera. We use white light illumination produced by the halogen bulb without any spectral filters. The PPM can adjust the ellipticity ε of polychromatic polarized illumination beam in a range from +45º (right-handed circular polarization) to -45º (left-handed circular polarization). The ellipticity is controlled by a rotating polarizer. It is convenient to set the zero orientation of the polarizer, θ=0º, when ε=+45º. In this case, according to Eq.(1), the illumination beam has the right circular polarization, and the microscope will be in the extinction state. The specimen under investigation is placed between the right circular polarizer and the left circular analyzer [28]. The orthogonal orientation of the polarizer θ=90º generates an opposite left circular polarization of the beam, ε=-45º. In this case, the specimen is located between two left circular polarizers. The ellipticity ε of the spectral polarization ellipses depends linearly on the orientation $\theta$ of the polarizer:

$$\varepsilon = 45º - \theta, \quad if\ 0º \leq \theta \leq 90º$$
$$\varepsilon = \theta - 135º, \quad if\ 90º < \theta < 180º$$

Usually, the polarizer is set according to $\theta = 180 * \Gamma/550$, where $\Gamma$ is the sample's expected retardance. In this case, the colors are more vivid. Before loading the sample, the image is white-balanced through the camera's software to achieve a gray background image, in which the intensities of red, green, and blue color channels are approximately equal.

In a separate experiment, we find that it is convenient to use an LED spot light bulb with a tunable hue, for example, Philips Hue PAR16 White and Color Ambiance Bulb (UPC 046677456672). The tunable light source allows one to compensate a possible color shading in the image.

## 2. PPM versus conventional polarizing microscope

Below we compare the performance of a conventional polarizing microscope and a PPM microscope in the visualization of a weekly birefringent specimen, representing a planar nematic slab of retardance $\Gamma$ =30 nm. A thermotropic nematic LC, 4-butyl-4-heptyl-bicyclohexyl-4-carbononitrile (CCN47) (purchased from Nematel GmbH) is filled into flat cells made of two parallel glass plates with a gap 1 μm. The inner surfaces of the glass plates are coated with a unidirectionally rubbed polyimide PI2555 (HD MicroSystems) to set the planar alignment of the LC director. CCN47 is maintained in the nematic phase at 45 ℃; at this temperature, its birefringence is $\Delta n = n_e - n_o \approx 0.03$; thus $\Gamma$ =30 nm. Here, $n_e$ and $n_o$ are the extraordinary and ordinary refractive indices, respectively.

### *Birefringent cell observed under a conventional polarizing microscope*

The cell is rotated in steps of 20º at the stage of a conventional polarizing microscope with crossed polarizer and analyzer. The cell's textures show maximum brightness when the director, which is also the optic axis makes the angles 45º and 135º with the polarizer, Figure 2. Note that the orientations of the director that differ by 90º show the same intensity of light transmission and could not be distinguished under a conventional microscope with crossed polarizers. The texture is extinct at orientations 0º and 90º. The 30-nm retardance cell is colorless at any orientation, as it should be, according to the Michel-Levy birefringence chart [2, 23]. According to the chart, the specimen should appear gray until $\Gamma$ reaches 250 nm. The brightness of the texture grows quadratically with $\Gamma$ for small $\Gamma < 50\ nm$. A non-birefringent specimen with zero retardance is extinct between crossed polarizers.

### *Birefringent cell observed under a PPM microscope*

The polarizer is set at an angle θ=10º. The textures of the $\Gamma$ =30 nm cell show colors that vary as the cell is rotated. However, the brightness of the texture does not change. Moreover, the image brightness in PPM mode is five times higher than the maximum brightness in the conventional mode with crossed polarizer and analyzer. This allows us to capture the PPM images with exposure times that are four times shorter than in a conventional approach. An important advantage of PPM over the conventional mode is that the mutually orthogonal patterns of the optic axes are readily distinguishable, as they are of different colors.

When the specimen is removed, the background turns gray and remains just as bright, which means that the user can also study non-birefringent structures that absorb or scatter light. If the specimen under investigation has both birefringent and non-birefringent regions, the image will be similar to a regular bright-field image with colored birefringent and gray non-birefringent areas. Of course, the non-birefringent structures should not be stained. If the specimen is stained, then the colors induced by birefringence could be distinguished from the colors caused by staining by using a differential PPM image, i.e., obtaining two images of complementary colors and extracting one from the other; the colors caused by staining will vanish, while the birefringence colors will double in intensity [29, 20]. One can also rotate the specimen, in which case the birefringent structures will change their color, while the permanent staining will preserve its color.

Let us summarize the advantages of a PPM microscope over a conventional polarization setup:
1. The user can deduce the orientation of the optic axis of birefringent specimens from their textures; there is no need to rotate the specimen.
2. The image in PPM is significantly brighter, and the user can employ a light source with lower intensity or/and increase the image acquisition speed.
3. The user can distinguish between the birefringent (anisotropic) and non-birefringent (isotropic) regions of the specimens.
4. The PPM mode of observation lifts the degeneracy of two patterns of the optical axis that are mutually perpendicular by rendering these patterns in different colors.

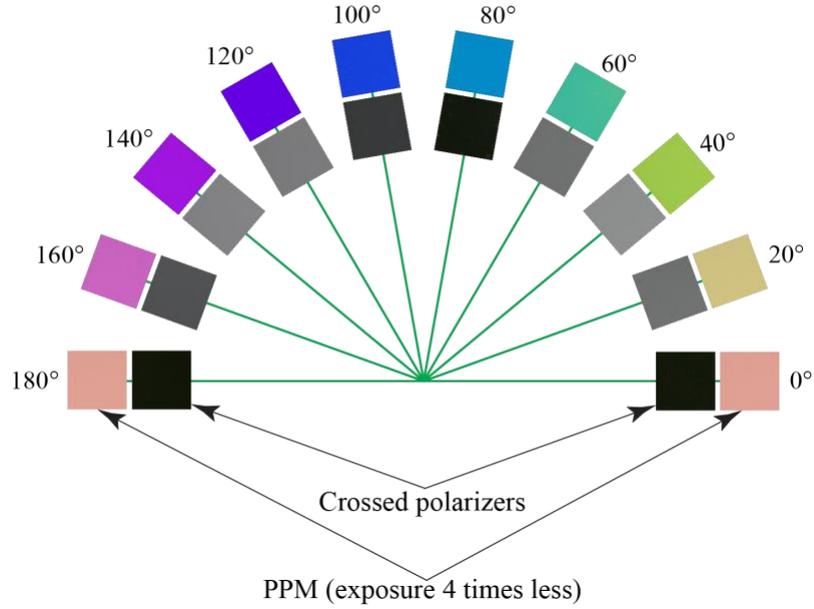

Fig. 2. Collage with textures of a 30-nm retardance cell in white light under a standard polarizing microscope with crossed polarizers (inner semi-circle) and under the same microscope equipped with a PPM (external semi-circle). The exposure time of PPM images was shorter by a factor of four.

### 3. Calibration of hue

As shown in Fig. 2, the hue changes continuously as the sample is rotated from 0° to 180°. Therefore, the hue value can be used to find the orientation of the slow axis of a birefringent sample. In the experiment illustrated by Fig.3, we study how the hue depends on the slow axis orientation. A first-order quarter waveplate Nikon P-CL with retardance 154 nm (measured at wavelength 535 nm) was used to calibrate the PPM. The test waveplate was placed at the microscope stage while the PPM was installed and rotated from 0° to 180° with 5° steps. The initial slow axis orientation was horizontal. The polarizer is set at the angle θ=50°. The experimental dependence of the hue on the orientation angle $\varphi$ of the slow axis is almost linear, Fig.3:

$$hue = \varphi / 180°.$$

In additional experiments, we find that the hue practically does not depend on retardance. Thus, $\varphi$ can be computed as:

$$\varphi = 180° \cdot hue.$$

According to the graph, the maximum error of the linear approximation is about 5° and occurs when φ≈70°, φ≈110°, and φ≈130°. In principle, instead of linear approximation, one can use spline interpolation to build a calibration curve or a calibration table to reduce the error. In particular, we did the spline interpolation using a Mathematica code to plot the director orientations in figures 4,5, and 6.

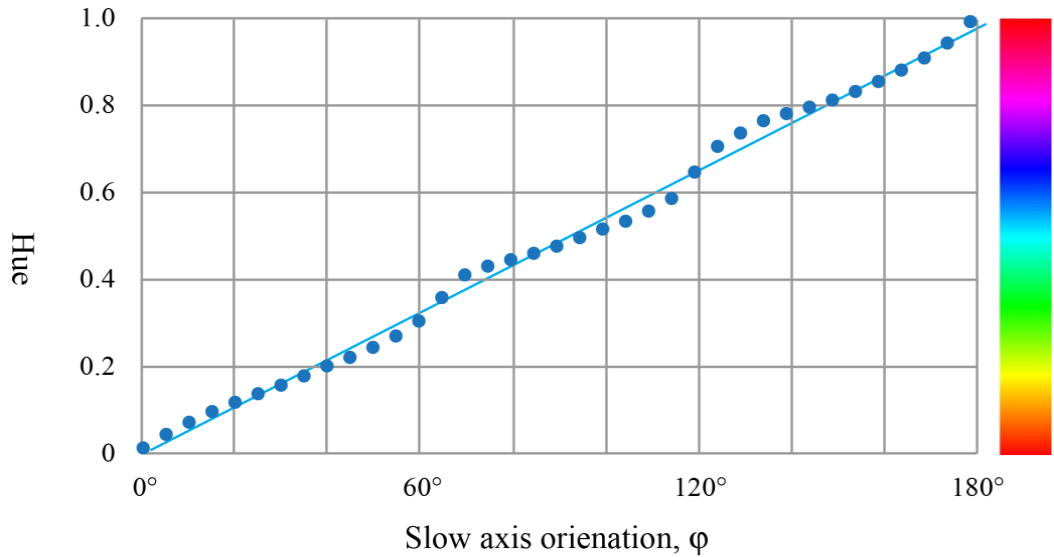

Fig. 3. Dependence of the hue on the slow axis orientation of a 154-nm retardance waveplate. The hue bar is shown on the right in color.

## 4. PPM versus MicroImager

The live image created by PPM is similar to the computed retardance/ orientation map generated by an LC-polscope (CRi, Waltham, MA) and a MicroImager (Hinds Instruments, Hillsboro, OR). The retardance/orientation map is constructed in the HSB (hue, saturation, brightness) color model, where the hue value depends on the slow axis (director) orientation, and the brightness is linearly proportional to the retardance. In particular, an LC-polscope utilizes the linear dependence of hue on orientation, while MicroImager applies a special conversion, as shown in Fig. 4. At the present time, production of LC-polscopes has been discontinued, but it is still used by many laboratories. The MicroImager is currently available on the market.

Figure 5 compares the images of a photopatterned nematic cell taken with PPM (top row) and MicroImager (bottom row). The MicroImager is a standalone device with a built-in monochromatic digital camera, 10x objective lens, and field of view 1140 µm x 1140 µm. The Olympus microscope BX51 is equipped with PPM and color digital camera Basler PowerPack acA1920-155um, 10x objective lens Olympus Ach and a field of view 2340 µm x 1460 µm. The PPM image was scaled and cropped to match the field of view and magnification of MicroImager. The color of PPM image was not modified. To match the color scales of both the MicroImager and the PPM, the zero orientation of PPM is moved to the blue end of the spectrum by 50º, as illustrated in the PPM color bar (top row in Fig.4). The bottom conversion color bar in Fig.4 shows that the dependence of hue on the director orientation is not linear in MicroImager. Therefore, the live PPM image (top of Fig.4) and computed MicroImager retardance/ orientation map (bottom) are slightly different.

The middle column in Fig.4 displays the hue distribution in PPM image and the grayscale orientation map, computed by MicroImager. Using the linear approximation described above in Section 3, we computed the director orientation distribution from PPM hue image. The difference between the PPM and MicroImager orientation maps is depicted on the right. The differential image has a light semi-circle because the magnification and adjustment do not match perfectly, which is especially noticeable at the red-blue transition. The callout displays an image area enhanced in terms of light intensity by 10 times. Within this area, the average difference between the optic axis orientation produced by the PPM and MicroImager is 3.1º with a standard deviation 1.2º.

Figure 4 confirms that the PPM and MicroImager director orientation maps match quite well. The discrepancy can be caused by the inaccuracy of the linear approximation, calibration of devices, scaling and aligning errors, and a slightly different focus. Currently, a PPM cannot measure the retardance amount, which is possible with MicroImager. However, the PPM generates the map of director orientation instantaneously (within the time needed by the camera to capture the image), while the MicroImager requires 7 sec. The PPM can utilize objective lenses (Pol or DIC) with any magnification, while MicroImager works with only four low-magnification objective lenses, 2x, 5x, 10, and 20x. The cost of PPM is considerably lower than that of the MicroImager. A PPM is actually a small mechanical attachment to an existing microscope. The MicroImager is a standing-alone microscope with custom software.

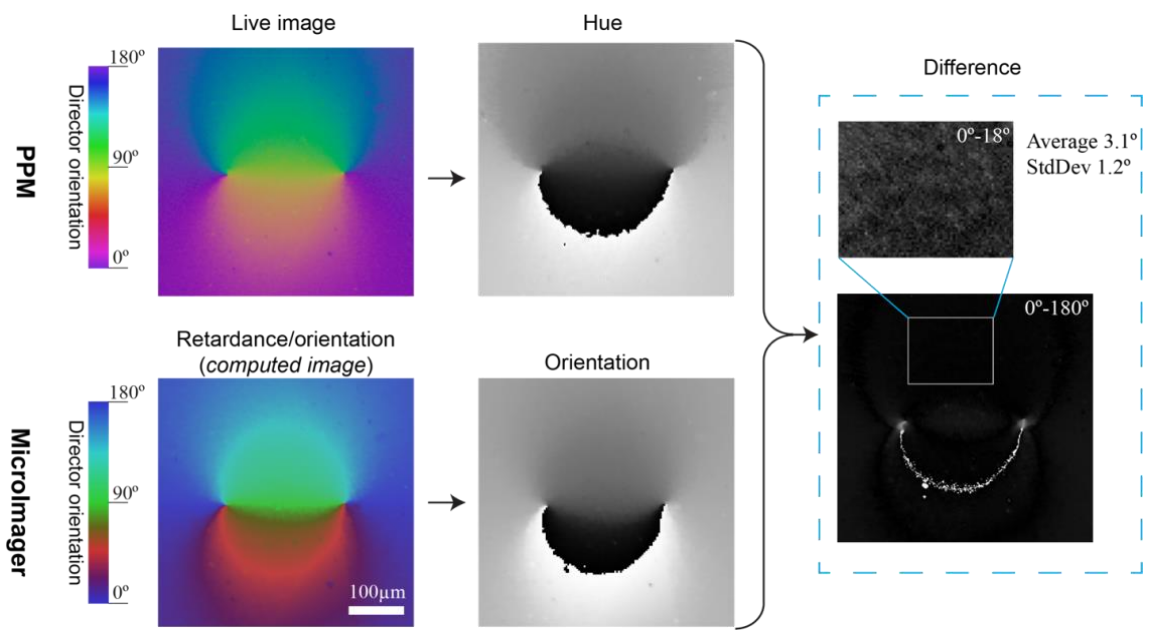

Fig. 4. Comparison of images of photopatterned nematic CCN47 cell taken with a PPM (top row) and a MicroImager (bottom row).

## 5. Examples of PPM application

Below we demonstrate the utility of PPM microscopy in the studies of liquid crystals in static and dynamic settings.

*Visualization of complex director patterns*

Complex director patterns were created by the technique of plasmonic surface photoalignment [30]. Glass plates, the inner surfaces of which are coated with an azo-dye Brilliant Yellow (BY) (Sigma-Aldrich), are assembled into sandwich-like cells with a gap thickness 3-5 μm. The BY layers are photoaligned using plasmonic metamasks with a pattern of elongated nanoslits. An unpolarized light beam passing through the mask acquires linear polarization along the direction perpendicular to the slit. The molecules of azo-dye, being irradiated by the locally polarized light, reorient along the normal to the light polarization. The resulting pattern of azo-dye orientation reproduces the pattern of nanoslits. Once the cell is filled with the nematic CCN47, its director follows the surface pattern of the two identical layers of azo-dye.

Figure 5 demonstrates that the PPM microscopy successfully maps the patterned 2D director field of CCN47 cells. The calibration data of figure 3 are used to plot the overlaying dashed profile representing the optic axis/director orientation. Figures 5a and b present the director field around pairs of topological defects of strength $m = \pm 1/2$ and $m = \pm 1$, respectively [31]. The PPM texture in Figure 5b resolves the narrow splitting of the $m = \pm 1$ defect cores into pairs of $m = \pm 1/2$ cores. The splitting is caused by a quadratic dependence of the elastic energy of a defect on its strength $m$, $F \propto m^2$. The equilibrium separation between two half-integer defects is set by the balance of the elastic and surface anchoring forces [32,33]. The effect provides means to measure the surface anchoring of photopatterned LCs [32,33]. Other complex patterns, such as periodic variation of the director splay and bend can also be mapped by PPM, Fig. 5c,d. The photopatterned director fields in Fig. 5c,d are used in applications such as optics [34], transport of colloids [35-37], bacterial and active flow control [38, 39], fabrication of liquid crystal elastomer coatings with dynamic topography [41], etc. Therefore, imaging the patterns details is of importance. The limitation of the PPM method is the same as that of other polarizing microscopy modes: the director twist along the light propagation direction changes the polarization state, which renders quantitative observations difficult. Another limitation is the range of recorded retardances which should be in the range 0 nm-250 nm. However, the most important advantage is that the PPM mode maps the director pattern over the entire field of view in a single short exposure, which allows one to explore dynamic processes, as described below.

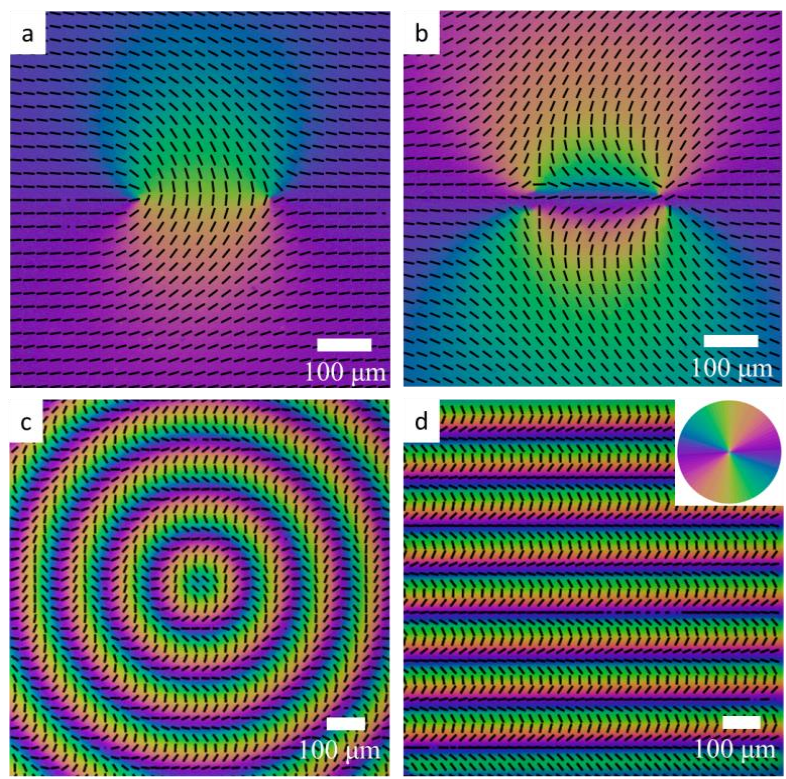

Fig. 5. Director orientation of photopatterned nematic CCN47 cells mapped using a PPM. a) A pair of topological point defects with strengths $m = +\frac{1}{2}$ and $m = -\frac{1}{2}$. Cell thickness, $d = 3.8 \ \mu m$. b) A pair of topological point defects with strengths $m = +1$ and $m = -1$. $d = 3.8 \ \mu m$. c) Circular pattern of splay and bend of the director, alternating along the radial directions. $d = 3.9 \ \mu m$. d) One-dimensionally periodic splay and bend of the director alternating along the vertical direction. $d = 4.6 \ \mu m$. In all experiments, $T = 45 \ °C$.

### *Visualization of dynamic director patterns*

In this experiment, we explore the shear-induced flows of a nematic lyotropic chromonic liquid crystal (LCLC), representing a 14 wt% dispersion of disodium cromoglycate (DSCG) (Spectrum Chemicals) in deionized water. The LCLC is sheared between two parallel disks (one of which is rotating) of the Linkam Optical Shearing System CSS450. The gap between the two plates is 10 μm, and the temperature is kept at 24 °C. The optical birefringence of the DSCG mixture at this temperature is negative, $\Delta n \approx -0.016$ [40].

Dynamics of the director field under shear is readily visualized by the PPM, Fig. 6 and 7. At low shear rates $\dot{\gamma} < 1$, the director aligns perpendicularly to the shear direction, along the vorticity direction, showing the so-called log-rolling behavior, Fig. 6 [40]. In the range $1 \text{ s}^{-1} < \dot{\gamma} < 500 \text{ s}^{-1}$, the director first becomes highly distorted and then reorients parallel to the shearing direction, Fig. 6. At high shear rates $\dot{\gamma} > 500 \text{ s}^{-1}$, the director becomes predominantly parallel to the shearing direction with periodic right and left tilts toward the vorticity axis, Fig. 6. The dynamic director patterns are similar to those established previously by Baza et. al. [40].

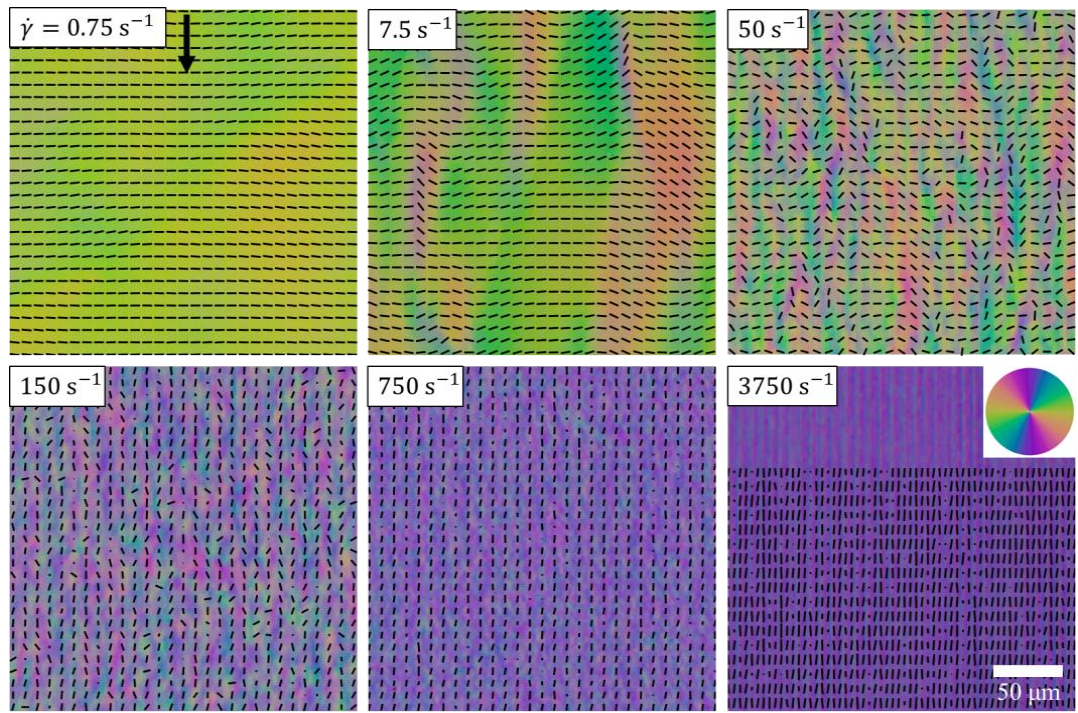

Fig. 6. PPM maps the director orientation of DSCG under different shearing rates. The arrow points along the shearing direction. $d = 10 \text{ μm}, T = 24°C$. In the panel for 3750 s$^{-1}$ the top stripe is intentionally left without the director pattern to demonstrate the periodic nature of distortions.

The PPM allows one to record the director pattern dynamics with a high temporal resolution. Figure 7 shows the time evolution of the director pattern under steady shear, $\dot{\gamma} = 22.5 \text{ s}^{-1}$, captured by PPM with 0.1 s steps. The speed of texture acquisition in a PPM is limited only by the camera's speed, a clear advantage over the techniques such as LC Abrio PolScope and Microimager, which require multiple exposures and thus a few seconds for the textural analysis.

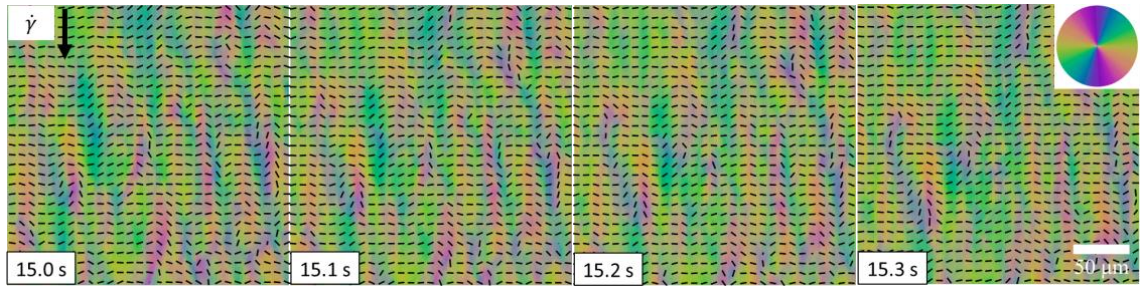

Fig. 7. Time evolution of the director field of a sheared DSCG mapped by a PPM. The arrow points along the shearing direction. $\dot{\gamma} = 22.5 \text{ s}^{-1}, d = 10 \text{ }\mu\text{m}, T = 24°\text{C}$.

## 6. Conclusions

Polychromatic polarizing microscope is a semiquantitative tool, which allows one to map the patterns of the optical axis in a single-shot exposure, thus enabling a fast temporal resolution. In combination with a pulsed light source, the new microscope makes it possible to visualize dynamic phenomena in liquid crystals and other birefringent materials. Pulse illumination in a stroboscopic mode might reveal additional details of repeatable dynamic processes.

**Acknowledgments**

The work is supported by NSF grant DMS-2106675. MS gratefully acknowledges funding from the NIGMS/NIH under Grant Nr. R01GM101701 and from the Inoue Endowment Fund.